\documentstyle[aps,twocolumn,floats,prl]{revtex}
\begin{document}
\title{Enhanced electron-phonon coupling at the Mo and W (110) surfaces
induced by adsorbed hydrogen}

\author{Bernd Kohler, Paolo Ruggerone, and Matthias Scheffler}
\address{
Fritz-Haber-Institut der Max-Planck-Gesellschaft,
Faradayweg 4-6, D-14\,195 Berlin-Dahlem, Germany}

\author{Erio Tosatti}
\address{
Instituto Nazionale di Fisica della Materia (INFM),
International School for Advanced Studies (SISSA),\\
and International Centre for Theoretical Physics (ICTP),
Miramare, I-34014 Trieste, Italy}
\date{Received 15 June 1995}
\maketitle
\begin{abstract}
The possible occurrence of either a charge-density-wave or a Kohn anomaly
is governed by the
presence of Fermi-surface nesting and the subtle
interaction of electrons and phonons.
Recent experimental and theoretical investigations
suggest such an effect for the
hydrogen covered Mo and W (110) surfaces. Using density-functional theory
we examine the electronic structure and the electron-phonon coupling of
these systems.
Besides  good agreement with the experimental phonon frequencies our
study provides a characterization and quantitative analysis
of an unusual scenario determining the electronic, vibrational, and
structural properties of these surfaces.
\end{abstract}
PACS numbers: 63.20.Kr, 68.35.Ja, 73.20.At, 73.20.Mf \\[0.1cm]

It is well established that the vibrational spectra of some
transition-metal {\em crystals} exhibit more or less pronounced anomalous
properties.
For example, in the phonon band structure of bulk Mo
a clear dip occurs at the ${H}$-point~\cite{powe68,zare83}.
Several theoretical studies~\cite{ho82,ho84,chen85,fu83,sing91}
support the idea that these anomalies are directly
related to a nesting of the Fermi surface~\cite{varmXX}.

If the electron-phonon ({\em e-ph}) coupling and the Kohn
anomaly become strong, an instability will arise,
leading to a Peierls-type, or more generally to a charge-density-wave
(CDW) type of insulator. The new ground state then contains
a frozen-in periodic lattice distortion. The distortion lowers
the kinetic energy of electrons near the Fermi surface
through an energy gap, which acts as the CDW order parameter.
In systems of effectively reduced dimensionality, such as layer and
chain bulk compounds, and crystal surfaces, one expects
nesting, and thus the occurrence of either strong Kohn anomalies,
or of a proper CDW ground state, to be generally enhanced~\cite{tos75}.
Nonetheless, with the exception of W(100) and Mo(100)~\cite{tos95},
no well-established cases of either giant Kohn anomalies {\em or} of
surface CDW's have yet emerged on real crystal surfaces.

The discovery of extremely sharp and deep indentations in the surface
phonon spectra of H/Mo(110) and H/W(110) at full hydrogen
coverage~\cite{hulp92a,lued94} has therefore come as
a surprise, raising much interest, and
many questions. The anomalies are seen at an incommensurate
{\bf k}-point, ${\bf Q}_{c1}$, along the [001]
direction (${\overline{\Gamma H}}$)
and, somewhat less visibly but
also unmistakably, at the commensurate
zone boundary point ${\bf Q}_{c2} = {\overline S}$ of the surface Brillouin
zone. Out of the
ordinary Rayleigh mode, {\em two} simultaneous anomalies appear to develop
at these two {\bf k}-points.
One, $\omega_1$, is extremely deep and sharp, and is only seen by
helium atom scattering (HAS).
The other, $\omega_2$, is instead soft and mild, and is observed by
both HAS and high resolution electron energy loss spectroscopy
(HREELS)~\cite{bald94}.
The agreement between HAS and HREELS dispersion curves of the
mild anomaly $\omega_2$ supports the idea that this indentation is
associated with the Rayleigh wave.
The interpretation of the deep and sharp anomaly
is less straightforward.
A close link between the surface phonon anomalies and
hydrogen vibrations~\cite{bald94} seems to be ruled out,
since the HAS spectra
remain practically unchanged when deuterium is adsorbed
instead of hydrogen~\cite{hulp92a,lued94}.

Recent density-functional-theory calculations of the Fermi surface
of the  hydrogen-covered Mo(110) $(1\times 1)$ surface
identified a hydrogen adsorption
induced quasi-one-dimensional Fermi-surface nesting in two different
directions~\cite{kohl95,rugg95}.
The calculated nesting vectors ${\bf Q}_{\rm c1}^{\rm th}$ along
$\overline{\Gamma {H}}$ and ${\bf Q}_{\rm c2}^{\rm th}$ along
$\overline{\Gamma S}$ are in excellent agreement with the
critical wave vectors at which HAS detected the anomalous behavior.

These experimental and theoretical results leave {\em two}
possible scenarios
still open for the ground state of these surfaces, namely:
\begin{enumerate}
\item[A.] The surface {\em e-ph} coupling is very strong,
the Ray\-leigh-wave
frequency at $2{\bf k}_{\rm F}$ becomes {\em imaginary}, and
the surface is
unstable.
A stabilizing distortion takes place and the true ground state
of the surface
is a quasi-one-dimensional CDW,
incommensurate along $\overline{\Gamma {H}}$, and commensurate
along $\overline{\Gamma S}$. In this hypothesis, suggested in
Ref.~\cite{tos95}, the surface
phonon anomalies $\omega_1$ and $\omega_2$ are therefore identified
as phase and amplitude CDW modes, well studied in
one-dimensional cases~\cite{lee74,giul78}.

\item[B.] The {\em e-ph} coupling and the associated Kohn ano\-maly
$\omega_2$ are only moderate.
Therefore, the unreconstructed surface structure remains stable and the
electronic instability is
not removed.
One should find only a slight downward shift of the Rayleigh-wave
frequency which remains {\em real}.
Within this picture, which was put forward in Ref.~\cite{kohl95}, the
mode $\omega_2$ is predominantly phonon-like and the other
mode, $\omega_1$, is predominantly due to an electron-hole excitation.
\end{enumerate}
Each of the two possibilities would be interesting and
uncommon. In order to ascertain whether scenario A or scenario B
is actually  realized, a  microscopic calculation of
the relevant {\em e-ph}
coupling, and of the ensuing shifts of surface phonon frequencies,
is needed.
A frozen-phonon calculation at ${\bf Q}_{c1}$ is out of
question, due to incommensurability.
Luckily, point ${\overline S}$ is instead simply commensurate, and here
the calculations are feasible.

We performed density functional theory
calculations using the local-density approximation for the
exchange-correlation energy functional~\cite{cepe80,perd81}.
For the self-consistent solution of the Kohn-Sham equations
we employed the full-potential linearized augmented plane-wave
method~\cite{blah85}.
Our code enables the direct calculation of atomic forces~\cite{yu91} and,
with a damped Newton dynamics scheme, the efficient determination of the
relaxed atomic structure~\cite{kohl9X}.
The substrates are modeled by five and seven layer slabs
repeated periodically and separated by a vacuum region of
thickness equivalent
to four substrate layers.
For the potential the $(l,m)$ representation within each muffin tin
(MT) sphere is taken up to $l_{\rm max}= 3$ while the kinetic-energy
cutoff for the interstitial region is set to 64\,Ry.
We choose the plane-wave cutoff for the wavefunctions to be 12\,Ry
and employ a $(l,m)$ representation in the MTs with $l_{\rm max}=8$ for
them.
The MT radii for the W and Mo atoms are chosen to be 1.27\,\AA.
For the hydrogen this radius is set to 0.48\,\AA.
Fermi smearing with a broadening of $kT_{\rm el}=0.07$~eV
is used in order to stabilize self-consistency and ${\bf k}$
summation~\cite{neug92}.
In the case of the W surfaces the valence and semi-core (core)
electrons are treated scalar (fully) relativistically while
for Mo all electrons are treated non-relativistically.
The in-plane lattice constants 3.14\,\AA\ (W) and
3.13\,\AA\ (Mo) calculated without including zero-point vibrations
are in good agreement with the respective measured bulk lattice
parameters  (3.163\,\AA\ and 3.148\,\AA\ for W~\cite{shah71}
and Mo~\cite{kata79},
respectively) at room temperature.

The necessary first step of our study is to determine the atomic and
electronic structures of the clean and hydrogen-covered (110)
surfaces.
In Table \ref{Tstructure} we summarize the calculated relaxation
parameters for these systems.
It turns out that the results for W(110) are very similar to those
obtained for Mo(110).
The investigations shine light on a long-standing problem concerning
the structure of W(110) covered with more than half a monolayer of
hydrogen. For this system Estrup's group observed a symmetry change in the
low energy electron diffraction pattern
and proposed that this might be caused by a displacement of the
top layer W atoms along the $[\overline{1}10]$ direction~\cite{chun86}.
Similar studies for H/Mo(110) did not provide any evidence for
such a top-layer-shift reconstruction~\cite{altm87}.
For both adsorbate systems we performed several structure optimizations
starting from different trial configurations in order to check whether
such a reconstruction is energetically favorable.
It turns out that for both substrates the hydrogen atom relaxes into a
quasi-threefold position (indicated as H in Fig. 1 in
Ref.~\cite{kohl95,rugg95}), and that the adsorption
reduces  the
inward relaxation of the clean surface considerably.
Furthermore, for H/Mo(110)~\cite{kohl95,rugg95} as well as
H/W(110)~\cite{aps95} we
find
no evidence for a {\em pronounced} top-layer-shift reconstruction.
However,  $y_1$ is non-zero, but very small, for both adsorbate
systems, which
indicates an anisotropic vibration of the top layer parallel to
the substrate with a favorable direction along $[\overline{1}10]$.

The similarities between Mo(110) and W(110) continue when we compare the
electronic structure of the relaxed surface systems. For both systems
the H adsorption alters the surface potential and induces a shift
of the $(d_{3z^2-r^2},d_{xz})$ band to higher binding
energies~\cite{kohl95,rugg95,aps95}.
In this way the Fermi line associated with this band is moved away
from $\overline{\Gamma N}$ into a band gap of the surface projected band
structure, and the
respective states become true surface states.
For both adsorbate systems
the shifted $(d_{3z^2-r^2},d_{xz})$ band is characterized by a high
density of states at the Fermi level, and the new Fermi contour gives rise
to a nesting~\cite{kohl95,aps95} .
The magnitudes of the calculated and measured critical wavevectors along
$\overline{\Gamma S}$ and $\overline{\Gamma H}$ are listed in Table
\ref{Tnesting}.
The agreement between theory and experiment for both systems is excellent.

The experimental phonon spectra of the (110) surfaces of Mo and W show two
distinct reactions to the adsorption of hydrogen.
Along $\overline{\Gamma H}$ and $\overline{\Gamma S}$ a monolayer
of hydrogen
induces a
softening of the Rayleigh and (to a smaller amount) of the
longitudinal~\cite{bald95}
wave while a stiffening of these modes is observed along
$\overline{\Gamma N}$.
Our goal is to investigate both effects theoretically.

A frozen phonon calculation at $\overline{S}$ is particularly convenient,
since at the zone boundary the Rayleigh-wave polarization
is purely vertical, and the second layer is immobile by symmetry.
In fact, at $\overline{S}$ the Rayleigh wave is not embedded in
the projected bulk phonon bands and thus it is strongly localized at
the surface.
Moreover, we find that the hydrogen vibrations are of no significant
importance for the calculated phonon frequencies.

In order to make our frozen phonon studies computationally
feasible we use an enlarged surface unit cell together with
a five layer slab
and reduce the plane-wave cutoff for the wave functions to 10\,Ry
(the convergence of total energy differences was tested).
The geometries are defined by the five layer relaxation parameters
presented in Table \ref{Tstructure}.
One should mention that some relaxation parameters, i.e.,
$\Delta d_{ij}$, change considerably when we perform a calculation
with a slab of five instead of seven metal layers.
We note, however, that the values of the critical
wavevectors and the position of the hydrogen with respect to the
substrate surface are practically insensitive to the thickness of the slab.
This indicates that the physical properties we are interested in,
e.g. the nesting features, are well localized surface phenomena and
that the results of our five layer slab studies can be trusted.

The phonon frequency was evaluated using the harmonic term
of a fourth-order polynomial fit to the total energy changes.
The total energies were calculated for five displacements
between 0\% and 3.6\% of the (110) surface interlayer spacing.
One might expect that the calculated frequencies,
especially the one for the $\overline{S}$-point phonon,
are sensitive to the Brillouin-zone sampling.
For instance, for the study of the $(2 \times 1)$ reconstructions
of the diamond (111) surface Vanderbilt and Louie~\cite{vand84}
employed a {\bf k}-point mesh which becomes logarithmically denser
close to the zone boundary.
We performed test calculations using 16, 56, and 120 {\bf k}-points
in the surface Brillouin zone in order to test the convergence
of our results, and found that a uniform {\bf k}-point mesh
of 56 points is sufficient.
The calculated frequencies are collected in Table \ref{Tphonon}.
At the $\overline{N}$ point our results reproduce the experimentally
observed {\em stiffening} of the Rayleigh-wave frequency as hydrogen is
adsorbed.
This agreement encourages to go on to the study of the
$\overline{S}$-point Rayleigh wave.
Here we find that the stiffening effects are overcompensated because of a
strong interaction between the phonons and the electron states
at the Fermi level.
In our calculations this leads to a  moderate  {\em lowering} of the
phonon energy
by 5.5\,meV for Mo and 6.1\,meV for W, in good agreement with the
experimental results. This strongly supports the conclusion that scenario B
is the correct one.

We may extract the surface {\em e-ph} coupling at the ${\overline S}$
point
by calculating the splitting of the two Kohn-Sham
eigenvalues folded back at the point midway between $\overline{\Gamma}$
and
$\overline{S}$ as a function of the nuclear displacement.
We obtain the gain in electronic energy which arises from the
lattice distortion of the frozen phonon, and this quantity can be
related to
the
electron-phonon interaction at the surface.
The calculated splitting $2\Delta\epsilon$ is practically linear
in the nuclear
displacement and the coefficient of the linear relation is
$2.81 \pm 0.10$ eV/\AA~for Mo and $4.41 \pm 0.06$ eV/\AA\ for W.
The larger value for W reflects a more efficient coupling between
the electrons
and phonons and suggests that the anomaly in the phonon band
of W(110) should be more pronounced.
This is indeed confirmed by the HAS experiments, which find that
the deeper dip is wider on H/W(110) than on H/Mo(110).
Furthermore, Hulpke and L\"udecke~\cite{hulp92a,lued94} reported the
occurrence of small structures in the HAS diffraction pattern for H/W(110)
at the critical wavevector.
These features, which were not observed for the  Mo substrate, might
indicate the start up of the formation of a CDW at W(110).

Our calculations of the electron-phonon interaction
at the (110) surfaces of both Mo and W and their changes due to hydrogen
adsorption pinpoint the phonon character of
the small anomaly $\omega_2$ and  identify  the subtle interplay
between the
electronic structure and the vibrational spectra of the transition
metal
surfaces. The results clearly support the interpretation that the
small dip
observed by
HAS and HREELS is due to a Kohn anomaly, while the huge dip arises from an
electron-hole
excitation.
Thus, scenario B is operative. There are in fact two independent
experimental
hints which support our conclusion.
If picture A were correct, and $\omega_1$ a phason, that phason
should exist at an incommensurate point like ${\bf Q}_{c1}$, but
should be absent
at the zone-boundary commensurate point ${\bf Q}_{c2} = {\overline
S}$~\cite{lee74}. However,
data near ${\overline S}$ clearly indicate a lower branch at that point
too.
Only if scenario B holds, is a second branch expected at ${\overline S}$
due to an electron-hole excitation and a good Fermi-surface nesting.
Furthermore,
recent HREELS
experiments~\cite{bald95}
reveal a positive temperature dependence of the small indentation ($d
\omega_2/dT > 0$)
consistent with that of a Kohn anomaly, i.e., a weakening of the
anomaly with increasing temperature.

In conclusion, we have discussed the atomic and electronic structures
of the clean and H-covered Mo(110) and W(110) surfaces
and studied the changes in the vibrational properties of these
surface systems upon hydrogen adsorption.
Two decisive experimental findings could be reproduced:
At the symmetry point $\overline{N}$ the Rayleigh phonon mode
is {\em stiffened} by the hydrogen adsorption, whereas for the
$\overline{S}$-point
Rayleigh wave we
obtain a {\em  weakening} of the phonon energy due to Fermi-surface
nesting.
The calculations support the interpretation~\cite{kohl95} that
scenario B is
in effect (the Kohn anomaly picture), as opposed to
the CDW picture~\cite{tos95} (scenario  A).
Thus, the modest anomaly seen in HAS and HREELS is a predominantly
phonon-like
excitation, and the giant indentation,
seen only in HAS, is predominantly due to an excitation of
electron-hole pairs.
The various elements identified in our analysis could be combined
in a future
study of the behavior of the H adatoms in order to analyze their
liquid-like
state features observed in the HREELS measurements~\cite{bald94}.

We thank
O.~Pankratov, S.~Wilke, M.~Balden, S.~de~Gironcoli, and C.~Bungaro
for stimulating discussions. Work at SISSA was supported by CNR, project
SUPALTEMP, and by the EU, contract ERBCHRXCT930342.

\begin{table}[tb]
\caption{Calculated relaxation parameters for the clean and H-covered
(110) surfaces of Mo and W. The height of the hydrogen above the
surface and its $[\overline{1}10]$ offset from the [001] bridge
position
are denoted by $d_{\rm H}$ and
$y_{\rm H}$, respectively.
The shift of the surface layer with respect to the substrate
is $y_1$.
The parameters $\Delta d_{ij}$ describe the percentage change of the
interlayer distance between the $i$-th and the $j$-th substrate layers
with
respect to the bulk interlayer spacing $d_0$.
For each system the results for a five layer (first line) and
for a seven layer (second line) slab are presented.}
\begin{tabular}{l|rrrrrr}
system
&$y_{\rm H}$
&$d_{\rm H}$
&$y_1$
&$\Delta d_{12}$
&$\Delta d_{23}$
&$\Delta d_{34}$        \\
&(\AA)
&(\AA)
&(\AA)
&($\%d_0$)
&($\%d_0$)
&($\%d_0$)              \\
\hline
Mo(110)         &$-$    &$-$    &$-$    &$-$5.9 &$-$0.8 &$-$    \\
                &$-$    &$-$    &$-$    &$-$4.5 &+0.5   &0.0    \\
\hline
H/Mo(110)       &0.63   &1.08   &0.05   &$-$2.7 &$-$0.4 &$-$    \\
                &0.60   &1.07   &0.03   &$-$2.1 &$+$0.1 &$-$0.1 \\
\hline
W(110)          &$-$    &$-$    &$-$    &$-$4.1 &$-$0.2 &$-$    \\
                &$-$    &$-$    &$-$    &$-$3.3 &$-$0.2 &$-$0.6 \\
\hline
H/W(110)        &0.68   &1.21   &$0.01$ &$-$1.3 &0.0    &$-$    \\
                &0.67   &1.20   &$0.05$ &$-$1.3 &$+$0.3 &+0.4   \\
\end{tabular}
\label{Tstructure}
\end{table}
\begin{table}[tb]
\caption{Theoretical Fermi surface nesting vectors compared to critical
wavevectors obtained by HAS and HREELS experiments
\protect{\cite{hulp92a,lued94,bald94}}.}
\begin{tabular}{llll}
direction       &system         &\multicolumn{2}{c}{
                                $|{\bf Q}_c|\ \ (\mbox{\AA}^{-1})$}\\
                                \cline{3-4}
                &               &theory &experiment\\
\hline  $\overline{\Gamma H}\ \ $
                        &H/Mo(110)      &0.86   &0.90 \\
                        &H/W(110)       &0.96   &0.95 \\
\hline  $\overline{\Gamma S}\ \ $
                        &H/Mo(110)      &1.23   &1.22\\
                        &H/W(110)       &1.22   &1.22\\
\end{tabular}
\label{Tnesting}
\end{table}
\begin{table}[tb]
\caption{Comparison of calculated frozen phonon energies and
experimental results values obtained by HAS
\protect{\cite{hulp92a,lued94}} and HREELS
\protect{\cite{bald94}}.
The numerical accuracy of the theoretical phonon energies
is about $\pm 1$\,meV.}
\begin{tabular}{llll}
phonon          &system         &\multicolumn{2}{c}{$E_{\rm ph}$\ \
(meV)}\\
                                \cline{3-4}
                &               &theory &experiment\\
\hline
$\overline{N}$  &W(110) &17.0   &14.5\\
                &H/W(110)       &19.8   &$\sim$19\\
\hline
$\overline{S}$  &Mo(110)        &23.7   &$\sim$21\\
                &H/Mo(110)      &18.2   &$<$16\\
                \cline{2-4}
                &W(110)         &19.1   &16.1\\
                &H/W(110)       &13.0   &11\\
\end{tabular}
\label{Tphonon}
\end{table}
\end{document}